\documentclass[pra,twocolumn,superscriptaddress,longbibliography]{revtex4-1}   
\usepackage{times,xspace}
\usepackage{amsbsy,amssymb,amsmath,bm,bbold} 
\usepackage{graphicx,color,epsfig,rotate} 
\usepackage{fancyhdr} 
\usepackage{epstopdf}
\usepackage{float}
\usepackage[colorlinks=true,linkcolor=blue,citecolor=blue,urlcolor=blue]{hyperref}

\usepackage{tikz}

\newcommand\encircle[1]{%
	\tikz[baseline=(X.base)] 
	\node (X) [draw, shape=circle, inner sep=1] {\strut #1};}

\def\bbbc{{\mathchoice {\setbox0=\hbox{$\displaystyle\rm C$}\hbox{\hbox 
to0pt{\kern0.4\wd0\vrule height0.9\ht0\hss}\box0}} 
{\setbox0=\hbox{$\textstyle\rm C$}\hbox{\hbox 
to0pt{\kern0.4\wd0\vrule height0.9\ht0\hss}\box0}} 
{\setbox0=\hbox{$\scriptstyle\rm C$}\hbox{\hbox 
to0pt{\kern0.4\wd0\vrule height0.9\ht0\hss}\box0}} 
{\setbox0=\hbox{$\scriptscriptstyle\rm C$}\hbox{\hbox 
to0pt{\kern0.4\wd0\vrule height0.9\ht0\hss}\box0}}}}

\DeclareMathAlphabet\mathbfcal{OMS}{cmsy}{b}{n}

\AtBeginDocument{%
	\newwrite\bibnotes
	\def\bibnotesext{Notes.bib}
	\immediate\openout\bibnotes=\jobname\bibnotesext
	\immediate\write\bibnotes{@CONTROL{REVTEX41Control}}
	\immediate\write\bibnotes{@CONTROL{%
			apsrev41Control,author="08",editor="1",pages="1",title="0",year="1"}}
	\if@filesw
	\immediate\write\@auxout{\string\citation{apsrev41Control}}%
	\fi
}%

\begin{document} 
\title{Correlated Nonreciprocity around Conjugate Exceptional Points}
\author{Arnab Laha}
\email{arnablaha777@gmail.com}
\affiliation{Institute of Spintronics and Quantum Information, Faculty of Physics, Adam Mickiewicz University, 61-614 Pozna\'n, Poland}
\affiliation{Department of Physics, Indian Institute of Technology Delhi, New Delhi-110016, India}
\author{Adam Miranowicz}
\affiliation{Institute of Spintronics and Quantum Information, Faculty of Physics, Adam Mickiewicz University, 61-614 Pozna\'n, Poland}
\author{R. K. Varshney}
\affiliation{Department of Physics, Indian Institute of Technology Delhi, New Delhi-110016, India}
\author{Somnath Ghosh}
\affiliation{Department of Physics, Indian Institute of Technology Jodhpur, Rajasthan-342037, India}

\begin{abstract}	
The occurrence of exceptional points (EPs) is a fascinating non-Hermitian feature of open systems. A level-repulsion phenomenon between two complex states of an open system can be realized by positioning an EP and its time-reversal ($\mathcal{T}$) conjugate pair in the underlying parameter space. Here, we report the fascinating nonreciprocal response of such two conjugate EPs by using a dual-mode planar waveguide system having two $\mathcal{T}$-symmetric active variants concerning the transverse gain-loss profiles. We specifically reveal a comprehensive all-optical scheme to achieve correlative nonreciprocal light dynamics by using the reverse chirality of two dynamically encircled conjugate EPs in the presence of local nonlinearity. A specific nonreciprocal correlation between two designed $\mathcal{T}$-symmetric waveguide variants is established in terms of their unidirectional transfer of light with a precise selection of modes. Here, the unconventional reverse chiral properties of two conjugate EPs allow the nonreciprocal transmission of two selective modes in the opposite directions of the underlying waveguide variants. An explicit dependence of the nonlinearity level on a significant enhancement of the nonreciprocity in terms of an isolation ratio is explored by investigating the effects of both local Kerr-type and saturable nonlinearities (considered separately). The physical insights and implications of harnessing the properties of conjugate EPs in nonlinear optical systems can enable the growth and development of a versatile platform for building nonreciprocal components and devices.
\end{abstract}  
\maketitle %

\section{Introduction}

The synergy of non-Hermitian quantum physics and photonics has been revealing a novel and promising direction for building a range of photonics components and devices \cite{Midya18}. An extensive study on the perturbation theory in quantum mechanics once revealed the occurrence of exceptional point (EP) singularities as an explicit mathematical feature of non-Hermitian or open systems \cite{Kato}. EPs usually appear as topological defects in the system's parameter space, affecting the eigenspace dimensionality, which results in the simultaneous coalescence of at least two coupled eigenvalues and the associated eigenstates \cite{Kato,Heiss00PRE,Berry2004,Heiss12JPA,Eleuch16PRA}. The parity-time ($\mathcal{PT}$)-symmetric systems (a special class of non-Hermitian system with real eigenvalues) \cite{Bender1998,Bender2007} encounter an EP at a spontaneous transition from real (exact-$\mathcal{PT}$-phase) to complex (broken-$\mathcal{PT}$-phase) eigenvalues \cite{Guo2009,Ruter2010,Ganainy18,Ozdemir19}. Recently, the engineering of ubiquitous non-Hermitian components (e.g. loss and gain) in photonic systems has revealed such EP-like mathematical objects as a powerful tool to manipulate and detect the energy-states of light \cite{Ganainy18,Miri19,Ozdemir19,Bergholtz2021,Wang23review,Yan23review,Parto21review}. A controlled variation of the system's parameters in the vicinity of EPs can immensely boost a versatile range of quantum-photonic technologies in the context of, e.g., asymmetric energy transfer \cite{Xu16}, programmable state-switching \cite{Arkhipov2023,Laha21EP4}, phonon lasing \cite{Zhang2018lasing}, coherent perfect absorption \cite{Wang21CPA}, slow-light engineering \cite{Goldzak18}, enhanced energy harvesting \cite{Lucas2021}, parametric instability \cite{Zyablovsky16} and highly-precise sensing \cite{Chen17sensor,Wiersig20}.

The concept of the occurrence of conjugate EPs has recently been introduced based on the complex parameter dependence of a non-Hermitian Hamiltonian \cite{Laha22}. This can be described by considering a generic two-level (without loss of generality for higher-order situations) non-Hermitian Hamiltonian $\mathcal{H}(\lambda)$, which depends on a complex parameter $\lambda=\lambda^{\text{R}}+i\lambda^{\text{I}}$. The associated eigenvalues $\mathcal{E}_{1,2}(\lambda)$ and the eigenvectors $\Psi_{1,2}(\lambda)$ would be analytical functions in the complex-$\lambda$ plane except at a singularity $\lambda=\lambda_{s}$, known as an EP. Concerning the imaginary part of the dependent parameter $\lambda$ (i.e., $\lambda^{\text{I}}$), the considerations of $\lambda^{\text{I}}<0$ and $\lambda^{\text{I}}>0$ ideally define two complementary variants of $\mathcal{H}(\lambda)$. Such two complementary systems can be correlated based on time-reversal ($\mathcal{T}$)-symmetry. Here, two variants of $\mathcal{H}(\lambda)$ under $\mathcal{T}$-symmetry separately host two EPs in the complex $\lambda$-plane at $\lambda_{s}=\lambda_s^{\text{R}}+i\lambda_s^{\text{I}}$ and $\lambda_{s}^{*}=\lambda_s^{\text{R}}-i\lambda_s^{\text{I}}$ (say, EP and its conjugate EP*, respectively), which are in the complex conjugate relation. Such two correlated EPs in two $\mathcal{T}$-symmetric complementary systems can be called as conjugate EPs.      

Unconventional light guidance mechanism based on the chirality of EPs has extensively been studied, where a sufficiently slow length-dependent gain-loss dynamics along a closed 2D loop around an EP can steer the adiabatic and nonadiabatic conversions of modes \cite{Gilary13na,Milburn15na}. Here, even though the adiabaticity is maintained in the sense of the exchange of eigenvalues for a quasistatic gain-loss variation \cite{Dembowski04}, the associated eigenmodes fail to meet adiabaticity while propagating along the length, which results in the conversion of all the modes into different particular dominating modes, based on the device chirality (in terms of direction of light propagation) \cite{Doppler16,Zhang19EP3,Laha18,Dey20,Laha20}. Such a chirality-based asymmetric transfer of modes has recently been explored to reveal a distinct reverse-chiral behavior of a pair of conjugate EPs, while dynamically encircling them in two $\mathcal{T}$-symmetric active variants of a waveguide-based optical system  \cite{Laha22}.  

Moreover, the reciprocity of such a chiral light guidance process can be broken by introducing nonreciprocal elements, where the occurrence of an EP can considerably enhance nonreciprocity \cite{Thomas16,Choi17isolator}. Nonreciprocal devices, such as isolators and circulators, allow only one-way light transmission with an asymmetric scattering matrix, which is indispensable to minimize unwanted back-reflection and multi-path interference in photonic circuits \cite{Caloz18}. However, the common magneto-optical approaches (such as a Faraday rotator), mainly applied for bulky free-space devices, are usually inefficient in enabling a sufficient nonreciprocity for photonic circuits. Hence, there are growing demands to achieve high nonreciprocity on the chip-scale footprint, where the chiral response of an EP in nonlinear media can play a crucial role in meeting such demands. Recently, an EP-induced mode-selective isolation scheme has been revealed, where local nonlinearity has served as an efficient tool to enable all-optical nonreciprocity without using any magneto-optical effect \cite{Laha20}. In this context, the chiral response of two conjugate EPs in nonlinear media could have immense potential in developing correlative nonreciprocal devices with highly precise mode manipulation. Moreover, the recently developed non-Hermitian formalism of Liouvillian super operators \cite{Minganti2019,Minganti2020} can also be exploited for the quantum implementation of our waveguide-based classical analysis to explore the correlated features of conjugate quantum EPs.   

In this article, we comprehensively report the correlated nonreciprocal response of two $\mathcal{T}$-symmetric active variants of a gain-loss assisted dual-mode planar waveguide, operating near two conjugate EPs. Here, all-optical nonreciprocity is achieved with the introduction of local nonlinearity. We investigate the hosting of conjugate EPs in complementary gain-loss parameter planes based on Riemann surface connections associated with two quasi-guided modes. Besides establishing the reverse-chiral response concerning the asymmetric mode conversion process driven by dynamical parametric variation in the vicinity of two conjugate EPs, we exclusively investigate the asymmetric nonreciprocal waveguidence mechanism in the context of all-photonic isolation through two $\mathcal{T}$-symmetric waveguide variants. Here, a correlation in the nonreciprocal transmission of selective modes with an enhanced isolation ratio (say, IR) through two complementary waveguides is established. Moreover, a comparative study on the individual effect of local Kerr-type nonlinearity and saturable nonlinearity is reported by showing the possibility of enhancing the IR significantly.

\section{Results and Discussion}

\subsection{Designing two time-symmetric active waveguide variants}

We design a 2D planar step-index optical waveguide having the geometrical dimensions $w=20\lambda/\pi$ (width) and $l=l_m\times10^3$ (length) with $l_m=7.5\lambda/\pi$ (i.e., both the dimensions are considered in the unit of wavelength $\lambda$), where we set $\lambda=2\pi$ corresponding to a normalized wavenumber $k=1$ (in a dimensionless unit). The designed waveguide with a glass based core ($n_{\text{co}}=1.5$), surrounded by a silica based cladding ($n_{\text{clad}}=1.46$), is distributed in the $xz$-plane, where $x\in[-w/2, w/2]$ and $z\in[0, l]$ are the transverse and propagation axes, respectively. The real (background) refractive index profile is considered as
\begin{equation}
\text{Re}[n(x)]=\left\{ 
\begin{array}{l}
n_{\text{co}}\,\,\\
n_{\text{clad}}\,\,
\end{array}
\begin{array}{l}
\,:-w/6\le x\le w/6,\\
\,:w/6\le |x|\le w/2.
\end{array}\right.
\label{renx} 
\end{equation}
Based on the chosen dimensional parameters and Re$n(x)$- profile, the designed waveguide supports only two scalar modes: the fundamental mode $\Psi_{\text{F}}$ and the first higher order mode $\Psi_{\text{H}}$ (scalar modal analysis is valid in the presence of a small index difference between the core and cladding; $\Delta n=0.04$).
\begin{figure}[b!]
	\centering
	\includegraphics[width=8.5cm]{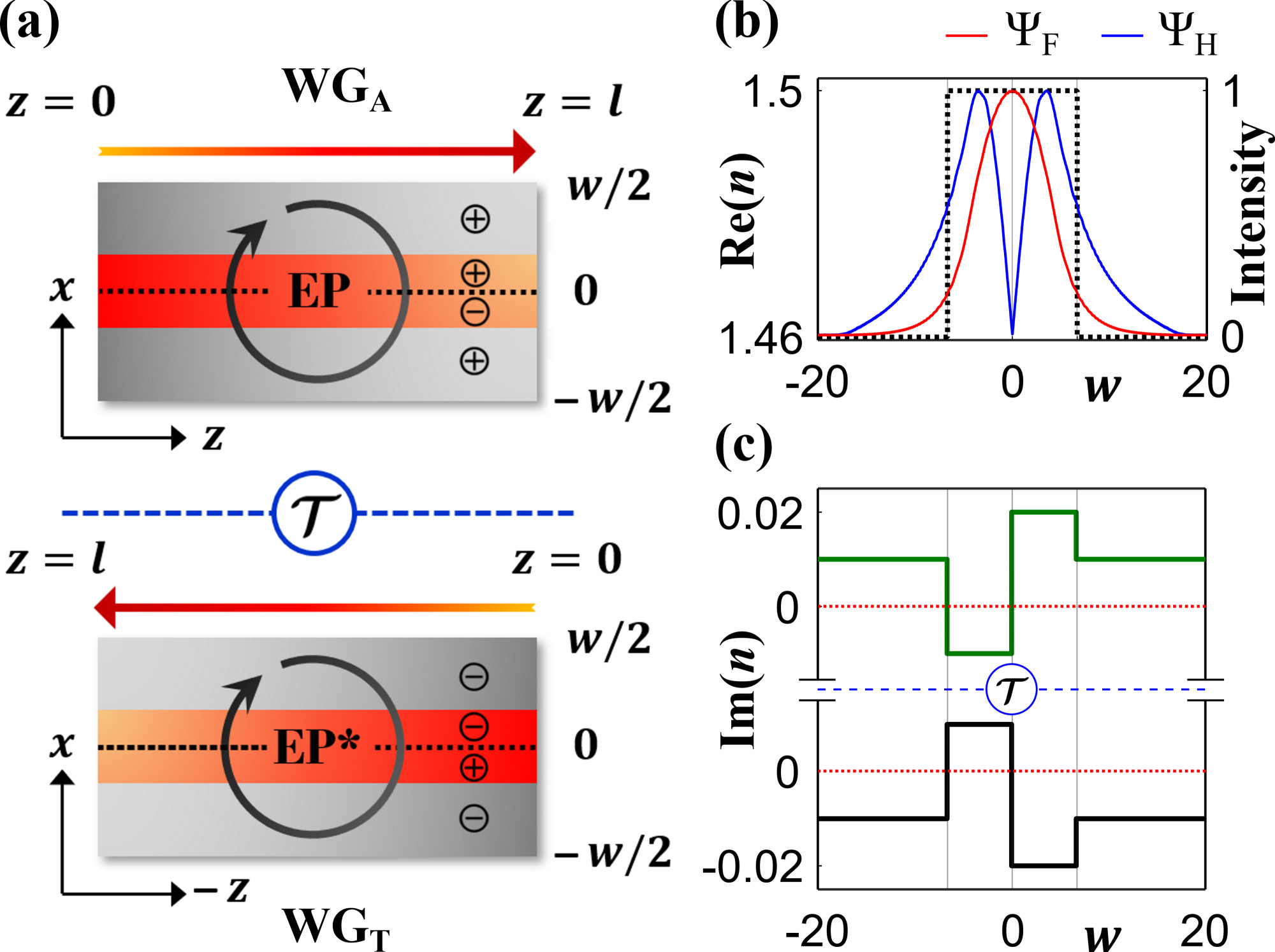}
	\caption{{\bf (a)} Schematic design of WG$_\text{A}$ and WG$_\text{T}$ ($\mathcal{T}$-symmetric) based on the framework of a gain-loss assisted planar waveguide. Two arrows indicate their opposite propagation directions. Circular plus and minus signs in different segments are associated with the positive (loss) and negative (gain) imaginary indices as in \eqref{imnx}. {\bf (b)} Transverse background refractive index profile, i.e., Re[$n(x)$], (dotted black line; corresponding to the left vertical axis) along with normalized intensity profiles of two supported modes $\Psi_{\text{F}}$ and $\Psi_{\text{H}}$ (corresponding to the right vertical axis). {\bf (c)} Transverse gain-loss distributions, i.e., Im[$n(x)$], for two $\mathcal{T}$-symmetric variants WG$_\text{A}$ (solid green line) and WG$_\text{T}$ (solid black line) for $\gamma=0.01$ and $\tau=2$.}
	\label{fig1}
\end{figure}
\begin{figure*}[t]
	\centering
	\includegraphics[width=\linewidth]{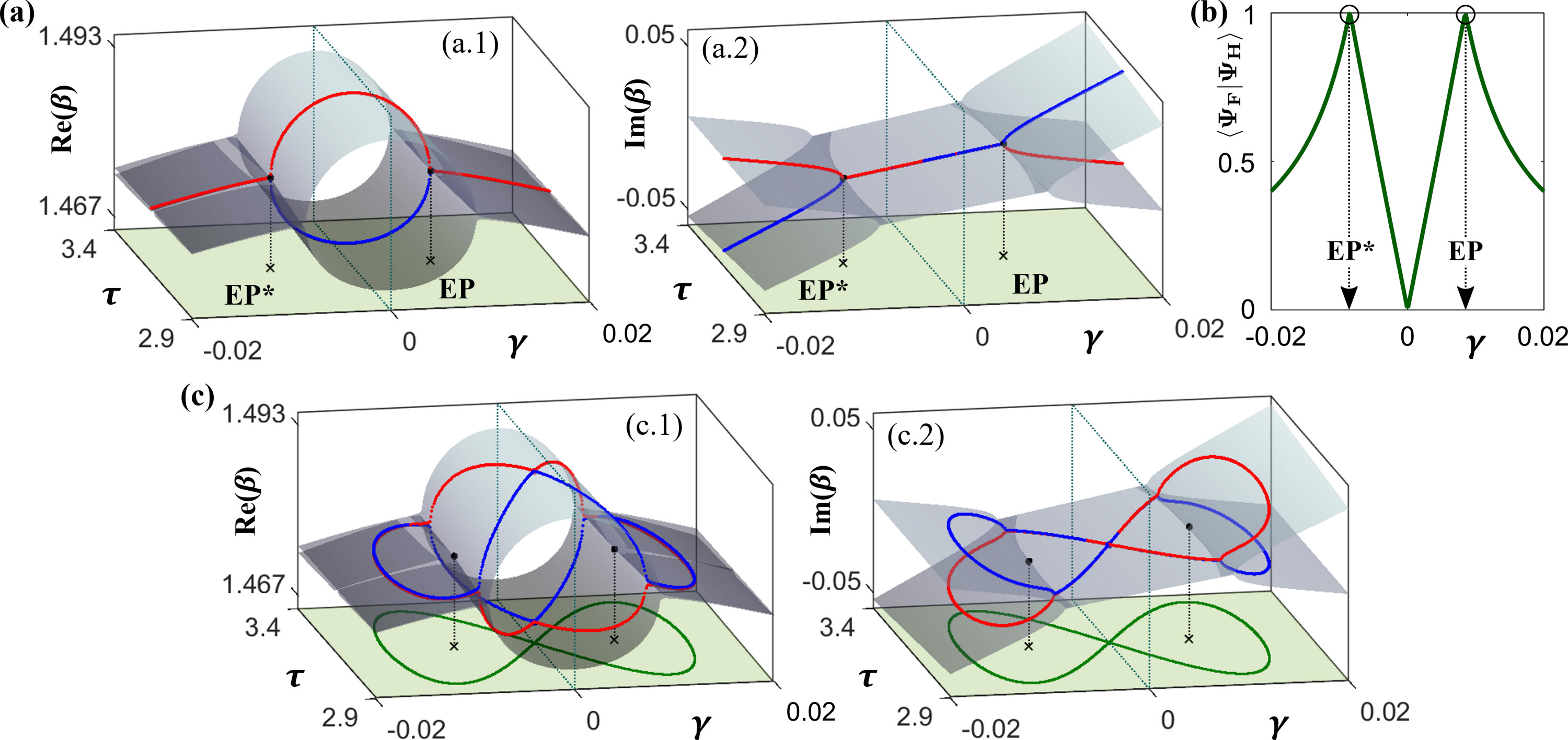}
	\caption{{\bf (a)} Connections between the Riemann surfaces associated with $\beta_{\text{F}}$ and $\beta_{\text{H}}$, while varying the control parameters $\gamma$ and $\tau$, simultaneously. (a.1) and (a.2) show the distributions of Re[$\beta$] and Im[$\beta$], respectively. Dotted red and blue curves represent the trajectories of $\beta_{\text{F}}$ and $\beta_{\text{H}}$ for a chosen $\tau=3.1607$, which reveal the encounter of two conjugate EPs based on the coalescence and bifurcations in Re[$\beta$] and Im[$\beta$] at $\gamma=\pm8.1\times10^{-3}$. Dotted blue squares separate the regions for WG$_\text{A}$ and WG$_\text{T}$. {\bf (b)} Variation of $\left\langle\Psi_{\text{F}}|\Psi_{\text{H}}\right\rangle$ with respect to $\gamma$ (when $\tau=3.1607$), which shows the coalescence of $\Psi_{\text{F}}$ and $\Psi_{\text{H}}$ via $\left\langle\Psi_{\text{F}}|\Psi_{\text{H}}\right\rangle=1$ at both EP and EP*. {\bf (b)} Parametric encirclement of two conjugate EPs in the ($\gamma$,$\tau$)-plane following \eqref{enc} (shown in the ground surfaces) and associated transfer process of $\beta_{\text{F}}$ and $\beta_{\text{H}}$ from their respective surfaces.}
	\label{fig2}
\end{figure*}

Now, we enable non-Hermiticity via the introduction of an unbalanced gain-loss profile [i.e., the imaginary part of $n(x)$] in the designed passive waveguide, which results in the coupling between two quasiguided modes $\Psi_{\text{F}}$ and $\Psi_{\text{H}}$. We can control such coupling with the modulation of a gain-loss profile in a 2D parameter space characterized by the gain-loss coefficient $\gamma$ and a loss-to-gain ratio $\tau$. Using this waveguide framework, we consider two complementary active variants connected via $\mathcal{T}$-symmetric Im[$n(x)$] profiles given by
\begin{equation}
\text{Im}[n(x)]=\left\{ 
\begin{array}{l}
-\,i\gamma\\+\,i\tau\gamma\\+\,i\gamma
\end{array}\right.
\begin{array}{c}
|\\\encircle{$\mathcal{T}$}\\|
\end{array}
\begin{array}{l}
+\,i\gamma\\-\,i\tau\gamma\\-\,i\gamma
\end{array}
\begin{array}{l}
\,:-w/6\le x\le 0,\\\,:0\le x\le w/6,\\\,:w/6\le |x|\le w/2.
\end{array}
\label{imnx} 
\end{equation}
Such two $\mathcal{T}$-symmetric waveguide variants, say WG$_\text{A}$ and WG$_\text{T}$ are shown in Fig. \ref{fig1}(a). Figure \ref{fig1}(b) shows the profile of Re$[n(x)]$ [dotted black line; given by \eqref{renx}] of the background framework along with the normalized intensity profile of two supported modes $\Psi_{\text{F}}$ and $\Psi_{\text{H}}$, whereas Fig. \ref{fig1}(c) shows the profiles of Im$[n(x)]$ of two active variants WG$_\text{A}$ and WG$_\text{T}$ (represented by green and black lines, respectively). As per the constraints of $\mathcal{T}$-symmetry, i.e.,  $\mathcal{T}:\left\{i,t,x\right\}\rightarrow\left\{-i,-t,x\right\}$ ($i$ is the imaginary quantity; $t$ and $x$ are the time and space coordinates, respectively), WG$_\text{A}$ and WG$_\text{T}$ host exactly two complex conjugate profiles of $n(x)$ with respect to the transverse axis [as can be understood from \eqref{imnx} and Fig. \ref{fig1}(c)]. Here, we have to consider two opposite propagation directions for WG$_\text{A}$ and WG$_\text{T}$ to maintain $\mathcal{T}$-symmetric equivalence based on the quantum-optical analogy $t\equiv z$.

\subsection{Riemann surface connections: Hosting conjugate EPs}

To host the pair of conjugate EPs, we study the interaction between two coupled eigenvalues associated with $\Psi_{\text{F}}$ and $\Psi_{\text{H}}$, while varying the control parameters $\gamma$ and $\tau$, simultaneously, within chosen ranges. In this context, an analytical treatment toward hosting conjugate EPs using a non-Hermitian Hamiltonian, which is analogous to our proposed waveguide-based system, is discussed in detail in an appendix. Here, the complex propagation constants ($\beta$-values), i.e., $\beta_{\text{F}}$ and $\beta_{\text{H}}$ (associated with $\Psi_{\text{F}}$ and $\Psi_{\text{H}}$, respectively) are considered as the system eigenvalues, which are calculated by computing the solutions of the 1D scalar wave equation $\left[\partial_x^2+k^2n^2(x)-\beta^2\right]\psi(x)=0$. We identify the connections between the Riemann sheets associated with coupled  $\beta_{\text{F}}$ and $\beta_{\text{H}}$ in Fig. \ref{fig2}(a) [with the distributions of Re($\beta$) and Im($\beta$) as shown in Figs. \ref{fig2}(a.1) and \ref{fig2}(a.2)], where the formation of a pair of conjugate EPs is clearly evident. Dotted red and blue curves show the trajectories of $\beta_{\text{F}}$ and $\beta_{\text{H}}$ concerning a continuous variation of $\gamma$, when we particularly choose $\tau=3.1607$. Here, we can observe a simultaneous bifurcation and a coalescence of the associated Re($\beta$) and Im($\beta$) values at $\gamma=-8.1\times10^{-3}$, as in Figs. \ref{fig2}(a.1) and \ref{fig2}(a.2), respectively. In contrary, a simultaneous coalescence and bifurcation of the associated Re($\beta$) and Im($\beta$) values can be observed at $\gamma=8.1\times10^{-3}$. Hence, two different circumstances corresponding to $\gamma<0$ and $\gamma>0$ for a specific $\tau$ refers to perfect complex conjugate situations (as the parameters $\gamma$ and $\tau$ are associated with Im[$n(x)$], i.e., gain-loss), which can ideally be observed in two active variants WG$_\text{A}$ and WG$_\text{T}$. The associated characteristics of $\beta_{\text{F}}$ and $\beta_{\text{H}}$ refer to the encounter of two conjugate EPs at ($\pm8.1\times10^{-3},3.1607$) (say, an EP and its conjugate EP* for WG$_\text{A}$ and WG$_\text{T}$, respectively) in the respective ($\gamma,\tau$)-planes. Topological dissimilarities in ARC-type interactions between $\beta_{\text{F}}$ and $\beta_{\text{H}}$ can clearly be observed alongside these conjugate EPs. The coalescence of the eigenmodes ($\Psi_{\text{F}}$ and $\Psi_{\text{H}}$) at both the conjugate EPs can be understood from the variation of $\left\langle\Psi_{\text{F}}|\Psi_{\text{H}}\right\rangle$ with $\gamma$ at a fixed $\tau=3.1607$, where $\left\langle\Psi_{\text{F}}|\Psi_{\text{H}}\right\rangle=1$ only at EP and EP*, as shown in Fig. \ref{fig2}(b).           

The effect of parametric encirclement of the embedded conjugate EPs in terms of chiral branch-point features is investigated in Fig. \ref{fig2}(c). Here, we consider two parametric loops in the 2D ($\gamma,\tau$)-plane according to the equations 
\begin{equation}
\gamma(\varphi)=\gamma_c\sin\left(\dfrac{\varphi}{2}\right)\quad\text{and}\quad \tau(\varphi)=\tau_c+r\sin(\varphi),
\label{enc}
\end{equation}
which leads to a closed and simultaneous variation of gain and loss around the EP and EP*. A slow variation of $\varphi\in[0,2\pi]$ governs the stroboscopic encirclements based on the characteristic parameters $\gamma_c$, $\tau_c$, and $r\in(0,1]$, where the conjugate EPs would be inside the parametric loop only for $|\gamma_c|>|\gamma_\text{EP}|$ ($\gamma_\text{EP}=8.1\times10^{-3}$; $\gamma$-value at the location of the EP). Here, the variations $\varphi:0\rightarrow2\pi$ and $\varphi:0\leftarrow2\pi$ enable a clockwise (CW) and a counter-clockwise (CCW) gain-loss variation around the EP for $\gamma_c>0$, and vice-versa around the EP* for $\gamma_c<0$. Such two parametric loops are shown in the ground surfaces of both Figs. \ref{fig2}(c.1) and \ref{fig2}(c.2) (for $r=0.3$, $\gamma_c=\pm1.5\times10^{-2}$. and $\tau_c=3.1607$), where the associated trajectories of coupled $\beta_{\text{F}}$ and $\beta_{\text{H}}$ are shown on their respective Riemann surfaces. Here, we observe that $\beta_{\text{F}}$ and $\beta_{\text{H}}$ are swapping their identities from their respective surfaces [concerning both Re($\beta$) and Im($\beta$), as can be seen in Figs. \ref{fig2}(c.1) and \ref{fig2}(c.2), respectively], and exchange their initial positions upon the completion of encirclement schemes. Such switching between complex $\beta_{\text{F}}$ and $\beta_{\text{H}}$ around both EP and EP* justify their branch-point behavior.

\subsection{Dynamically encircled conjugate EPs: Asymmetric transfer of modes in a linear medium}

\begin{figure}[t]
	\centering
	\includegraphics[width=8.5cm]{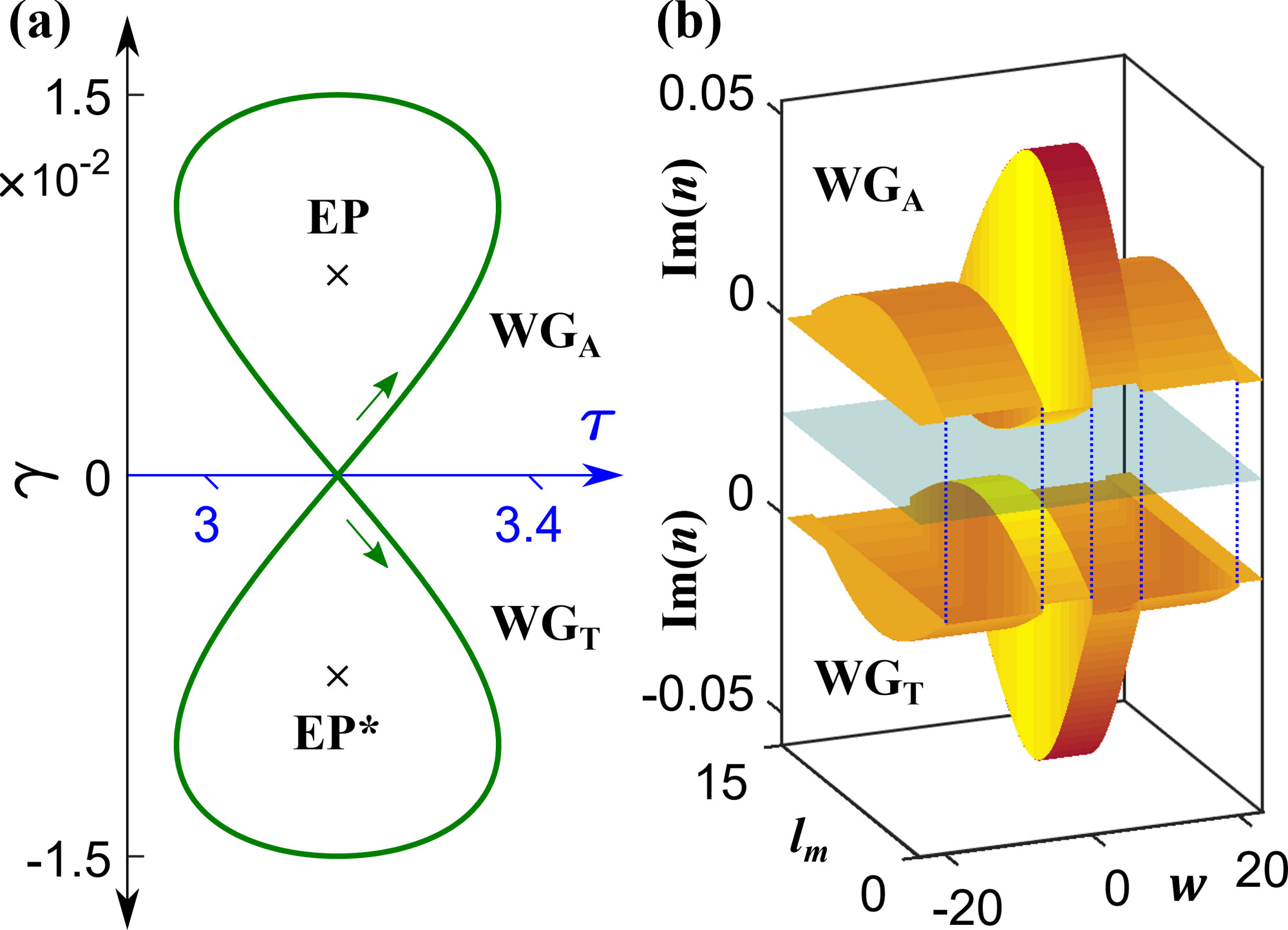}
	\caption{{\bf (a)} Parametric loops to encircle an EP and its conjugate EP* in the ($\gamma$,$\tau$)-plane [following \eqref{enc}]. {\bf (b)} Associated dynamical variation of gain-loss profiles, i.e., two complex conjugate active potentials (separated via a transparent plane), experienced by two $\mathcal{T}$-symmetric waveguide variants WG$_\text{A}$ and WG$_\text{T}$.}
	\label{fig3}
\end{figure}

Here, we consider the length dependence (analogous to the time dependence) on the encirclements of the conjugate EPs to study the correlative propagation dynamics of light (modes) through two $\mathcal{T}$-symmetric waveguide variants. Figure \ref{fig3}(a) shows the chosen parametric loops for WG$_\text{A}$ and WG$_\text{T}$ [to encircle EP and EP*; exactly the same loops, as can be seen in the ground surfaces of Fig. \ref{fig2}(c)]. We map the associated gain-loss distribution along the length ($z$-axis) of respective waveguides. Here, the reversal of the time axis ($t\rightarrow -t$) under the constraint of $\mathcal{T}$-symmetry allows us to consider mapping obligatorily in opposite directions (i.e., $z\rightarrow -z$ as  $t\equiv z$) for WG$_\text{A}$ and WG$_\text{T}$. Hence, we distribute the gain-loss profile from $z=0$ to $z=l$ based on the encirclement of EP (EP*) governed by $\varphi:0\rightarrow2\pi$ ($\varphi:2\pi\rightarrow0$) for WG$_\text{A}$ (WG$_\text{T}$). Such a $z$-dependent gain-loss distribution can be implemented by reconsidering \eqref{enc} as a function of $z$ as 
\begin{equation}
\gamma(z)=\gamma_c\sin\left(\dfrac{\pi z}{l}\right)\quad\text{and}\quad
\tau(z)=\tau_c+r\sin\left(\dfrac{2\pi z}{l}\right)
\label{denc}
\end{equation}
Figure \ref{fig3}(b) shows two complex conjugate 2D Im($n$)-profiles [governed by \eqref{denc}] to encircle the EP and EP* dynamically. Here, the CW and CCW directions of encirclements are realized through one complete pass of light in the forward direction ($z:0\rightarrow l$) and backward direction ($z:l\rightarrow 0$), respectively for WG$_\text{A}$, and vice-versa for WG$_\text{T}$.  

\begin{figure}
	\centering
	\includegraphics[width=8.5cm]{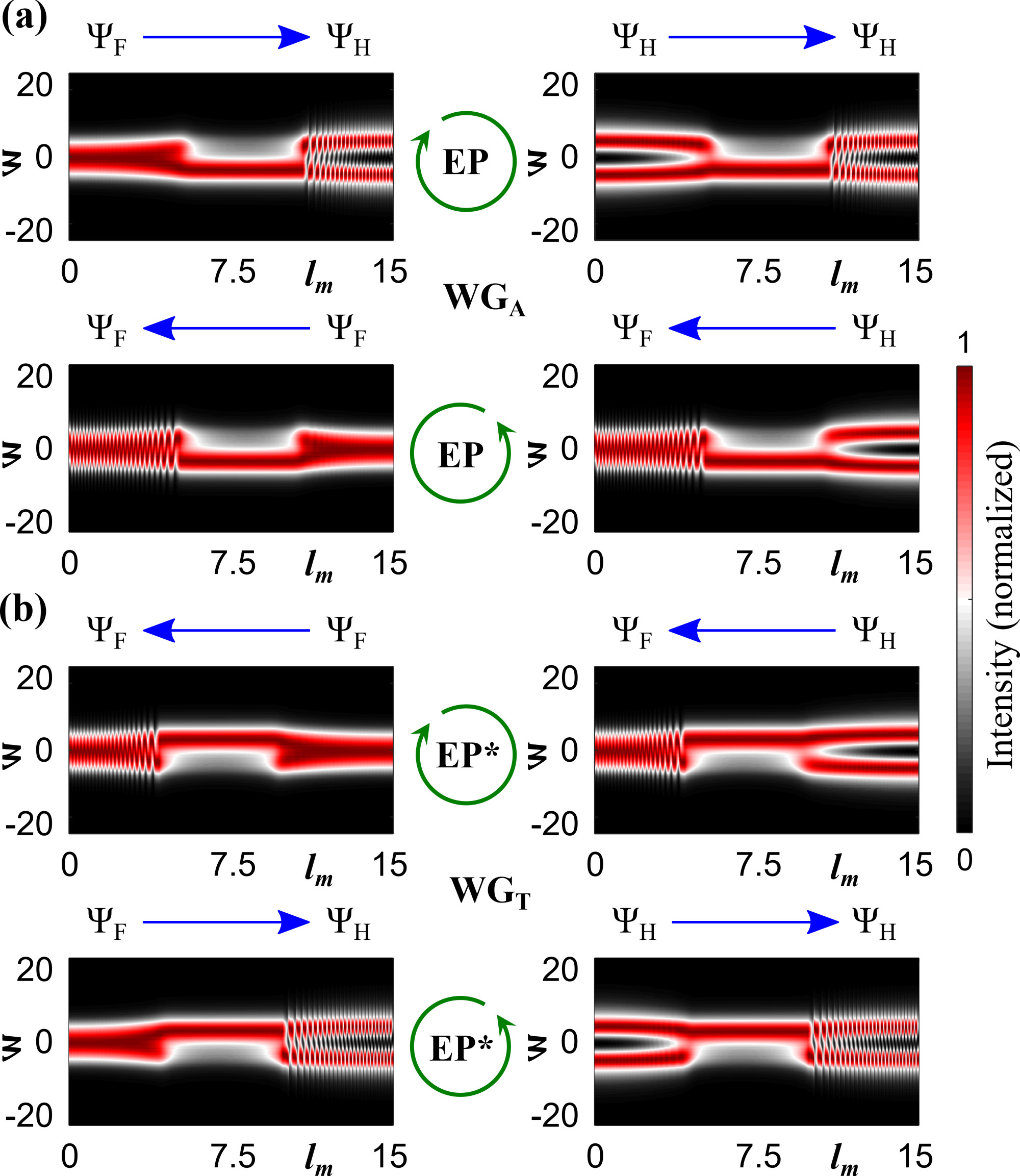}
	\caption{{\bf (a)} Propagation dynamics of $\Psi_{\text{F}}$ and $\Psi_{\text{H}}$ through WG$_\text{A}$ (upper panel) from $z=0$ to $z=l$ (associated with the CW dynamical EP-encirclement) followed by the asymmetric conversions $\{\Psi_{\text{F}},\Psi_{\text{H}}\}\rightarrow\Psi_{\text{H}}$; (lower panel) from $z=l$ to $z=0$ (associated with the CCW dynamical EP-encirclement) followed by the asymmetric conversions $\{\Psi_{\text{F}},\Psi_{\text{H}}\}\rightarrow\Psi_{\text{F}}$. {\bf (b)} Similar modal dynamics through WG$_\text{T}$ (upper panel) for the CW dynamical encirclement around the EP* with $z:l\rightarrow0$, exhibiting the asymmetric conversions $\{\Psi_{\text{F}},\Psi_{\text{H}}\}\rightarrow\Psi_{\text{F}}$; (lower panel) for the CCW dynamical encirclement around the EP* with $z:0\rightarrow l$, exhibiting the asymmetric conversions $\{\Psi_{\text{F}},\Psi_{\text{H}}\}\rightarrow\Psi_{\text{H}}$. Intensities are renormalized at each step of evolution along $z$ to show the inputs and outputs clearly.}
	\label{fig4}
\end{figure}

Now, we implement the scalar beam propagation method to investigate the individual light transmission through WG$_\text{A}$ and WG$_\text{T}$. Under the paraxial and adiabatic (which corresponds to a sufficiently slow variation of gain-loss along $z$-direction) approximation, we implement the scalar beam propagation equation, i.e.,
\begin{equation}
-2ik\partial_z\Psi(x,z)=\left[\partial_x^2+k^2\Delta n^2(x,z)\right]\Psi(x,z),
\end{equation}  
associated with both $\Psi_{\text{F}}$ and $\Psi_{\text{H}}$ with extremely fine split-step computation [with $\Delta n^2(x,z)\equiv n^2(x,z)-n_{\text{clad}}^2$]. Figure \ref{fig4} shows the resultant propagation characteristics, while considering the dynamical encirclements around the EP and EP* in WG$_\text{A}$ and WG$_\text{T}$, respectively. Here, we initially verify the linear response (i.e., without any nonlinearity) in the context of an asymmetric transfer between the modes, which occurs due to the failure of the adiabatic approximation led by a dynamically encircled EP, despite the associated omnipresent $\beta$-switching process. Here, the EP (or EP*) itself acts as a source of chirality, which mainly steers the response of the underlying system in the context of a direction-dependent transfer of modes. Such an unconventional modal dynamics can be observed for both WG$_\text{A}$ and WG$_\text{T}$, as shown in Figs. \ref{fig4}(a) and \ref{fig4}(b), respectively.

To enable a dynamical encirclement of the EP in the CW direction, we consider the propagation of light from $z=0$ to $z=l$ (forward direction) in WG$_\text{A}$. We can observe the corresponding dynamics of $\Psi_{\text{F}}$ and $\Psi_{\text{H}}$ in the upper panel of Fig. \ref{fig4}(a), where $\Psi_{\text{F}}$ is converted into $\Psi_{\text{H}}$, following the adiabatic expectation. However, $\Psi_{\text{H}}$ violets the system adiabaticity, i.e., it becomes restructured and remains as $\Psi_{\text{H}}$. Thus a light signal launched at $z=0$ of WG$_\text{A}$ is converted into a dominating $\Psi_{\text{H}}$ at $z=l$. The lower panel of Fig. \ref{fig4}(a) shows the modal transitions, while considering light propagation in the backward direction ($z:l\rightarrow0$; associated with the CCW encirclement process). Here, $\Psi_{\text{F}}$ dominates at the output $z=0$ with the asymmetric conversions $\left\{\Psi_{\text{F}},\Psi_{\text{H}}\right\}\rightarrow\Psi_{\text{F}}$, where only $\Psi_{\text{H}}$ maintains the adiabatic expectations (unlike the case for the CW encirclement process). Thus during the dynamical encirclement of an EP, the system partially maintains the adiabaticity, which however turns into a fascinating chiral light dynamics, where irrespective of the excited modes at the input, the device delivers two different dominating modes in the opposite directions.    

Such a violation in the system adiabaticity around an EP, can be predicted with the associated nonadiabatic correction terms ($\mathbb{N}_{\text{F}\rightarrow\text{H}}$ and $\mathbb{N}_{\text{H}\rightarrow\text{F}}$ for the adiabatic expectations $\Psi_{\text{F}}\rightarrow\Psi_{\text{H}}$ and $\Psi_{\text{H}}\rightarrow\Psi_{\text{F}}$, respectively) from the adiabatic theorem \cite{Gilary13na}. These corrections mainly rely on the accumulated relative-gain ($\Delta\gamma^{\text{ad}}_{\text{F,H}}$) factor during the transition of modes as (generalized with a quantum-optical analogy under the operating condition)
\begin{equation}
\mathbb{N}_{\text{F\{H\}}\rightarrow\text{H\{F\}}}\propto-\{+\}\exp\displaystyle\int_0^l\Delta\gamma^{\text{ad}}_{\text{F,H}}(\gamma,\tau)dz.
\label{ncf}
\end{equation} 
Here, $\Delta\gamma^{\text{ad}}_{\text{F,H}}$ can be estimated from the relative difference between the average loss ($\gamma^{\text{m}}$) accrued by the individual modes. The adiabatic trajectories of Im($\beta$)-values [as shown in Fig. \ref{fig2}(c.2)] for $\Psi_{\text{F}}$ and $\Psi_{\text{H}}$ gives the associated $\gamma^{\text{m}}$ with $\oint\{\text{Im}(\beta)/2\pi\}d\varphi$.

Here, the variant WG$_\text{A}$ operating with a dynamically encircled EP gives $\Delta\gamma^{\text{ad}}_{\text{F,H}}>0$ for the CW direction, whereas $\Delta\gamma^{\text{ad}}_{\text{F,H}}<0$ for the CCW direction. These particular relations result in the domination of the $\mathbb{N}$-factor associated with the amplifying exponent of $\Delta\gamma^{\text{ad}}_{\text{F,H}}$ over the overall adiabatic expectations, whereas cooperation of the $\mathbb{N}$-factor corresponding to the decaying exponent of $\Delta\gamma^{\text{ad}}_{\text{F,H}}$ with the adiabatic expectations. Hence, the domination of $\mathbb{N}_{\text{H}\rightarrow\text{F}}$ in the forward direction yields the nonadiabatic transition of $\Psi_{\text{H}}(\rightarrow\Psi_{\text{H}})$, whereas the cooperation of $\mathbb{N}_{\text{F}\rightarrow\text{H}}$ supports the adiabatic conversion of $\Psi_{\text{F}}(\rightarrow\Psi_{\text{H}})$. On the other hand, the domination of $\mathbb{N}_{\text{F}\rightarrow\text{H}}$ in the backward direction yields the nonadiabatic transition of $\Psi_{\text{F}}(\rightarrow\Psi_{\text{F}})$, whereas the cooperation of $\mathbb{N}_{\text{H}\rightarrow\text{F}}$ supports the adiabatic conversion of $\Psi_{\text{H}}(\rightarrow\Psi_{\text{F}})$. The detailed analytical predictions completely support our numerical beam-propagation results for WG$_\text{A}$, as shown in Fig. \ref{fig4}(a). From the dependence of the relative-gain factor $\Delta\gamma^{\text{ad}}_{\text{F,H}}$ on the EP-induced asymmetric mode conversions, one can generically conclude that the mode transiting with a lower average loss ($\gamma^{\text{m}}$) follows the adiabatic rules, whereas its coupled counterpart evolves nonadiabatically.

Now, if we consider the dynamical encirclement around EP*, then the concerned waveguide variant WG$_\text{T}$ exhibits reverse-chiral dynamics compared to the chiral behavior of WG$_\text{A}$, as can be seen in Fig. \ref{fig4}(b). During the encirclement in the CW directions, $\Psi_{\text{F}}$ and $\Psi_{\text{H}}$ transmit along the backward direction ($z:l\rightarrow0$) with $\Delta\gamma^{\text{ad}}_{\text{F,H}}<0$, which allows the nonadiabatic transfer of $\Psi_{\text{F}}$ and the adiabatic transfer of $\Psi_{\text{H}}$ with the asymmetric conversions $\left\{\Psi_{\text{F}},\Psi_{\text{H}}\right\}\rightarrow\Psi_{\text{F}}$ at $z=0$ [as shown in the upper panel of Fig. \ref{fig4}(b)]. In this case, $\Psi_{\text{H}}$ evolves with a lower $\gamma^m$ and maintains the adiabatic expectations. In contrary, the modal transmissions in the forward direction ($z:0\rightarrow l$) of WG$_\text{T}$ with a positive relative-gain factor ($\Delta\gamma^{\text{ad}}_{\text{F,H}}>0$) yields the delivery of the dominating $\Psi_{\text{F}}$ with the asymmetric conversions $\left\{\Psi_{\text{F}},\Psi_{\text{H}}\right\}\rightarrow\Psi_{\text{H}}$, while considering the encirclement in the CCW direction [as shown in the lower panel of Fig. \ref{fig4}(b)]. Here, $\Psi_{\text{F}}$ evolves with a lower $\gamma^m$ and maintains the adiabatic expectations. Hence, based on the constraints of the $\mathcal{T}$-symmetry, we exclusively demonstrate interesting opposite chiral responses of two active variants designed on the same background waveguide system, where the opposite encirclement directions around the EP and EP* result in the delivery of modes of the same order.

\subsection{Effect of nonlinearity on the asymmetric state-transfer process: Enabling nonreciprocity around two conjugate EPs}

\begin{figure}
	\centering
	\includegraphics[width=7.5cm]{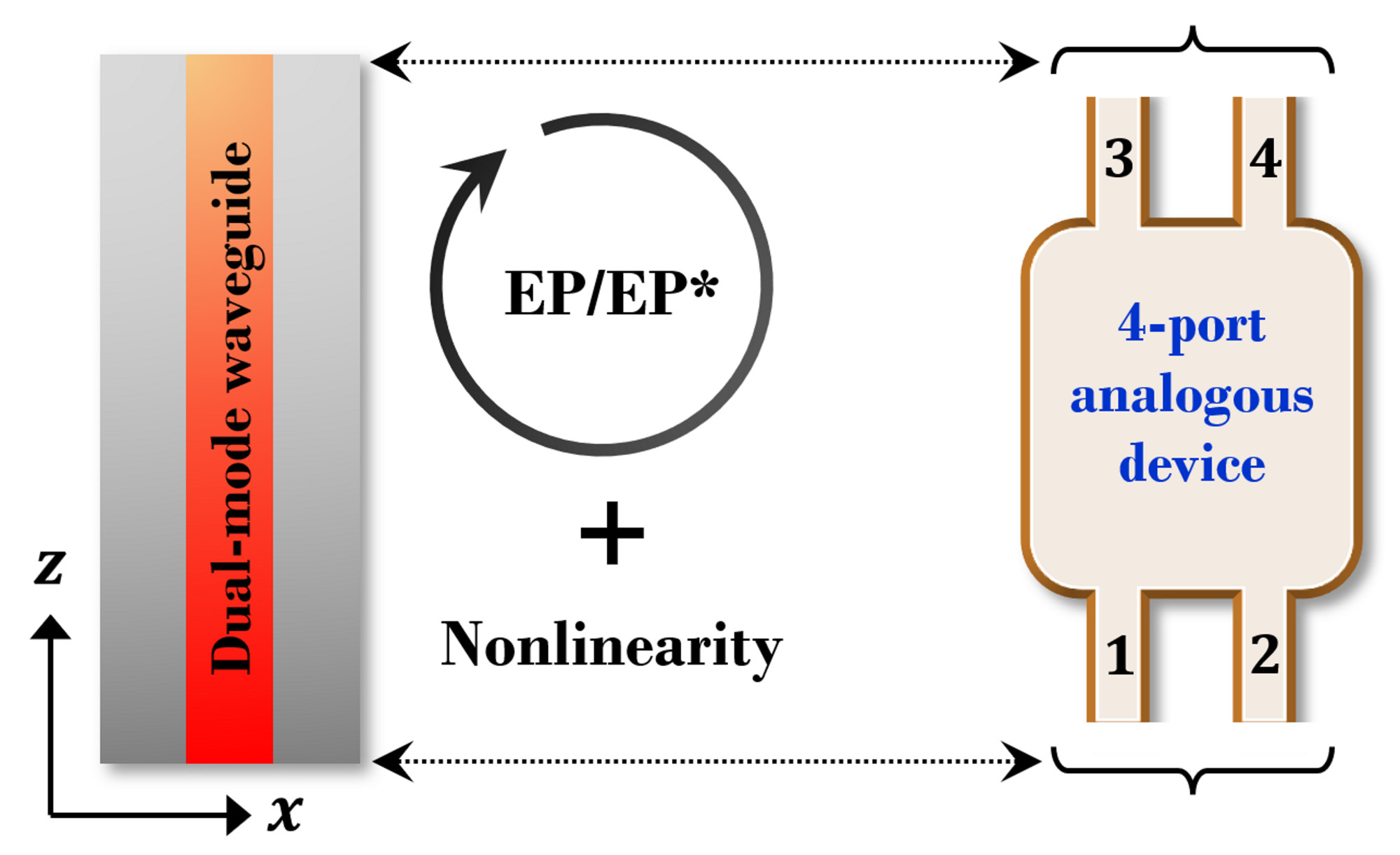}
	\caption{A schematic analogy between a 4-port optical device and our designed dual-mode waveguide operating with a dynamically encircled EP or EP* in the presence of nonlinearity. This analogy is essentially drawn to construct a $4\times4$ $S$-matrix [given by \eqref{smatrix}] considering all the possible transmissions.}
	\label{fig5}
\end{figure}

The direction-dependent light transmission process with the asymmetric transfer of modes in a dual-mode waveguide system (as described for two variants) can be mimicked by a scattering matrix ($S$-matrix) given by
\begin{equation}
[\Psi^m_{\text{op}}]_{4\times 1}=[S_{mn}]_{4\times 4}\left[\Psi^n_{\text{in}}\right]_{4\times 1}.
\label{smatrix}
\end{equation}
Equation (\ref{smatrix}) describes the operation of an analogous four-port device via $S$-matrix formalism, as shown in Fig. \ref{fig5}. Here the elements of $[S]$ can be calculated as $S_{mn}=\left\langle\Psi^n_{\text{in}}|\Psi^m_{\text{op}}\right\rangle$ with $\{m,n\}\in\{1,2,3,4\}$. We can safely consider the top-left and bottom-right blocks of $[S_{mn}]$ as $2\times2$ null matrices in order to neglect the possible reflections from the same port of the designed waveguide. Here, the forward ($T_{\text{F}}$) and backward ($T_{\text{B}}$) transmissions can be estimated with $T_{\text{F}}=|\max\left(\text{B}_{bl}\right)|^2$ and $T_{\text{F}}=|\max\left(\text{B}_{tr}\right)|^2$ (where, $\text{B}_{bl}$ and $\text{B}_{tr}$ represent the bottom-left and top-right blocks, respectively). Now, it can be understood that if $[S]$ defines the scattering matrix for WG$_\text{A}$, then the analytical transpose form of $[S]$ would be associated with WG$_\text{T}$ (the numerical values of matrix elements would indeed be different for WG$_\text{A}$ and WG$_\text{T}$ due to the presence of two opposite gain-loss profile). 

In the linear regime, the chirality-drove asymmetric mode conversion process in a particular waveguide variant follows Lorentz's reciprocity with a symmetric $S$-matrix, i.e., $[S]=[S]^T$. Now, the direction dependence on the system's response can bring up a special interest in achieving one-way transmission, which is compulsory for designing nonreciprocal devices. However, the presence of nonreciprocity obligatorily indicates the breakdown of Lorentz's reciprocity with an asymmetric $S$-matrix, i.e., $[S]\ne[S]^T$ \cite{Jalas13}. In this context, unidirectional transmission with a symmetric scattering matrix was reported in a photonic circuit \cite{Feng11unidirectional}, where isolation is not realizable \cite{Jalas13,Fan12comment}.

\begin{figure}[t]
	\centering
	\includegraphics[width=8.5cm]{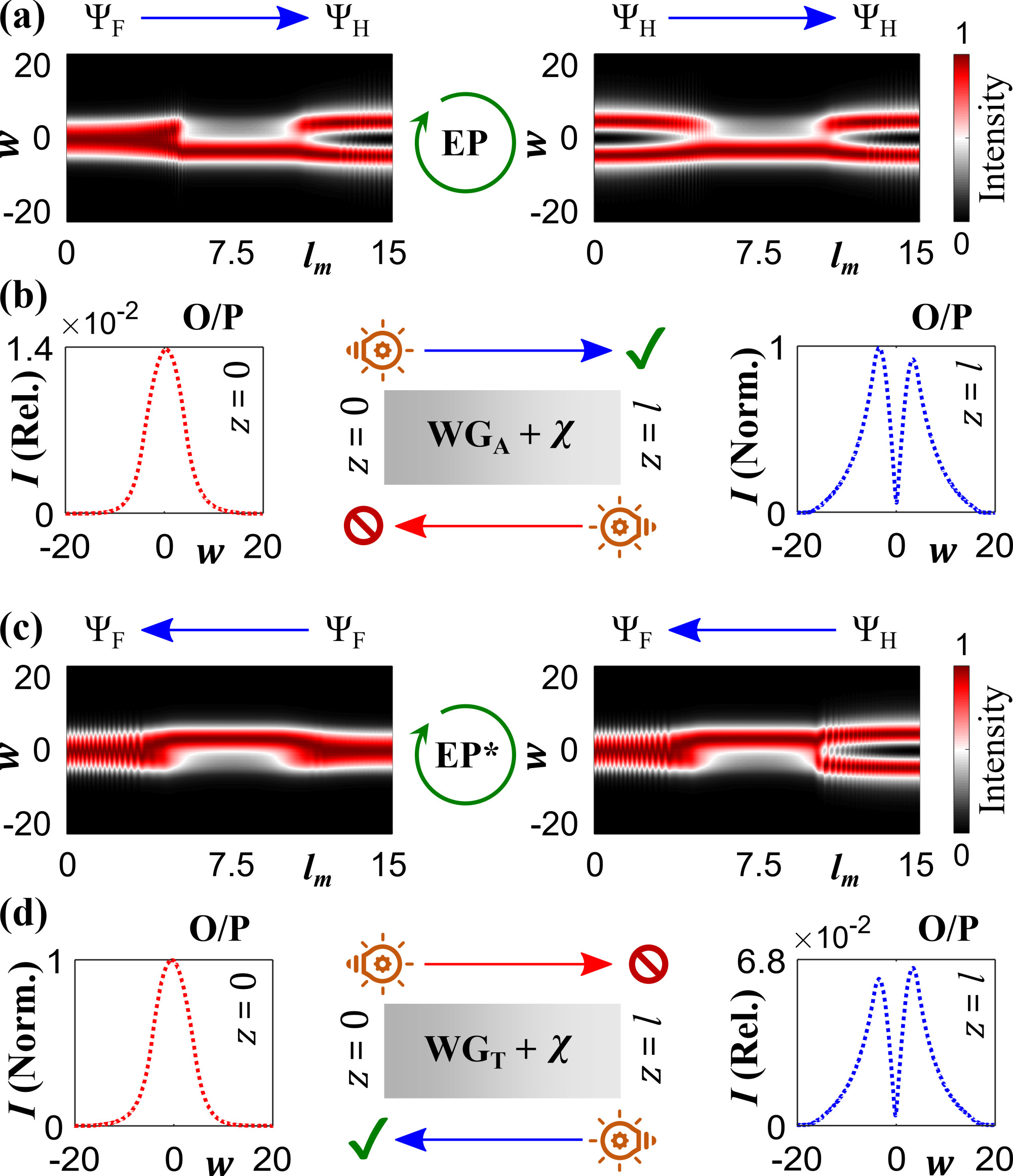}
	\caption{{\bf (a)} Nonreciprocal transition of modes with the asymmetric conversions $\{\Psi_{\text{F}},\Psi_{\text{H}}\}\rightarrow\Psi_{\text{H}}$ through WG$_\text{A}$ which is active in the forward direction ($z:0\rightarrow l$; associated with the CW dynamical EP encirclement process). {\bf (b)} Schematic nonreciprocal response of WG$_\text{A}$ (which allows light to pass in the forward direction, however, blocks in the backward directions) along with outputs (O/P) at $z=l$ (for the allowed path $z:0\rightarrow l$) and $z=0$ (for the blocked path $z:l\rightarrow 0$. {\bf (c)} Nonreciprocal transition of modes with the asymmetric conversions $\{\Psi_{\text{F}},\Psi_{\text{H}}\}\rightarrow\Psi_{\text{F}}$ through WG$_\text{T}$ which is active in the backward direction ($z:l\rightarrow 0$; associated with the CW dynamical encirclement of the EP*). {\bf (d)} Schematic nonreciprocal response of WG$_\text{T}$ (which allows light to pass in the backward direction, however, blocks in the forward directions) along with outputs (O/P) at $z=0$ (for the allowed path $z:l\rightarrow 0$) and $z=l$ (for the blocked path $z:0\rightarrow l$). For both WG$_\text{A}$ and WG$_\text{T}$, the self-normalized outputs are shown for their active directions, whereas relative outputs are shown in their blocked directions.}
	\label{fig6}
\end{figure}

\begin{figure}
	\centering
	\includegraphics[width=8.5cm]{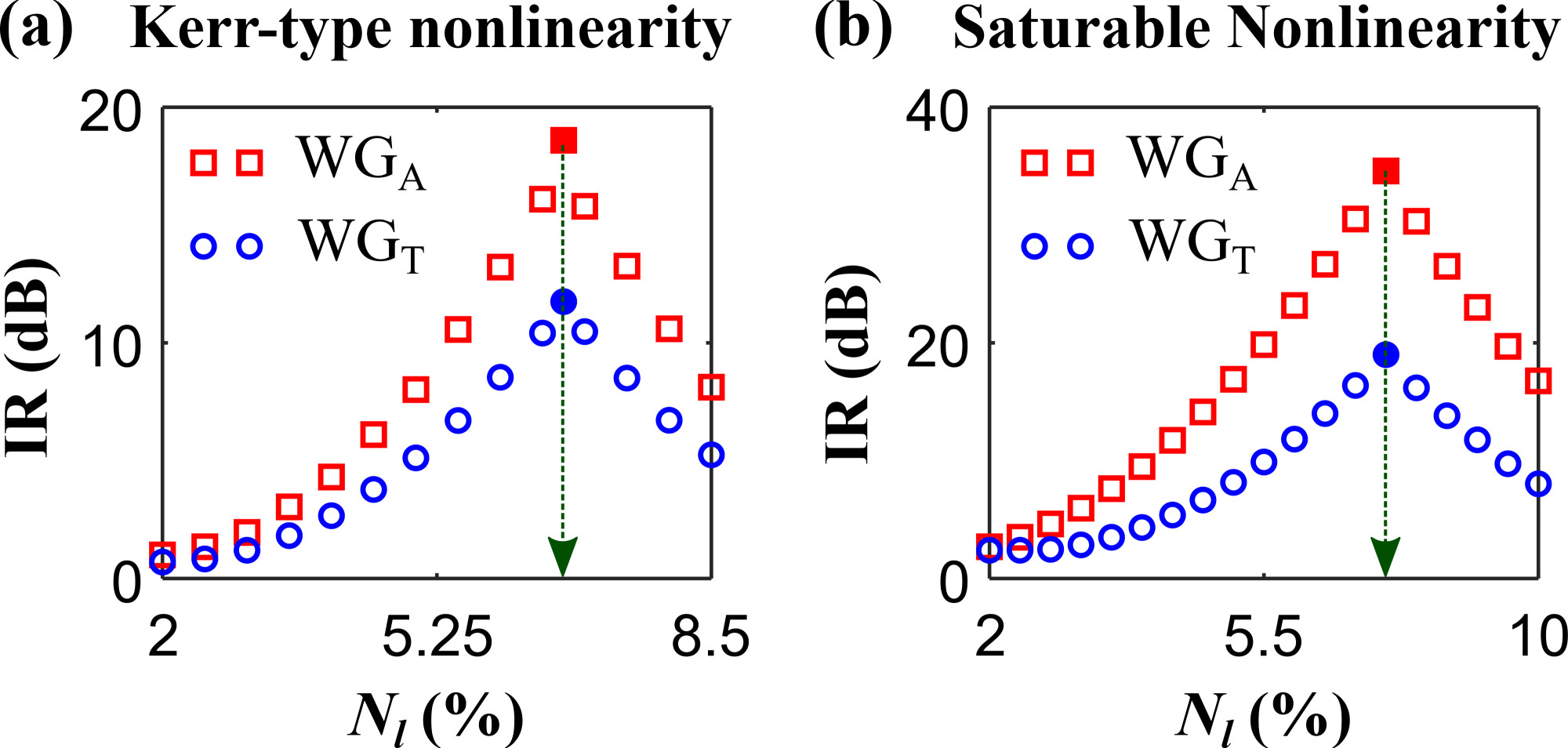}
	\caption{Dependence of the isolation ratio (IR) on the local nonlinearity level ($N_l$), while considering {\bf (a)} Kerr-type nonlinearity and {\bf (b)} saturable nonlinearity, separately. Red square and blue circular markers show such a variation of IR for WG$_\text{A}$ and WG$_\text{T}$, respectively. The green arrows in both (a) and (b) indicate the largest values of the IRs, as achieved at the same $N_l$.}
	\label{fig7}
\end{figure}

In order to break the reciprocity in EP-induced light dynamics, we exploit the effect of local nonlinearity. We schematically represent our proposed scheme in Fig. \ref{fig5} with an operational analogy between one of the designed dual-mode waveguide variants (hosting a dynamically encircled EP or EP*) with nonlinearity and a 4-port isolator device. Here, we quantify a particular nonlinearity level as $N_l=(\Delta n_{\text{NL}}/\Delta n)\times100\%$ (with $\Delta n=0.04$; for the designed passive waveguide), where the variation of $\Delta n_{\text{NL}}$ depends on the modal field-intensities ($I\equiv|\Psi|^2$) for a particular nonlinear coefficient ($n_2$). Here, we initially study the effect of Kerr-type nonlinearities to achieve an adequate level of nonreciprocity for both waveguide variants (in terms of an isolation ratio, say, IR) with proper optimization. Then, we also explore the effect of saturable nonlinearities to enhance the IR further and perform a quantitative comparison. We consider the forms of two different types of nonlinearities, viz., 
\begin{subequations}
	\begin{align}
	&\text{Kerr-type nonlinearity:}&\Delta n_{\text{NL}}(x,z)&=n_2I,\label{kerr}\\
	&\text{Saturable nonlinearity:}&\Delta n_{\text{NL}}(x,z)&=\dfrac{n_2I}{(1+I/I_s)}.\label{saturable}
	\end{align}
\end{subequations}%
$I_s$ in \eqref{saturable} defines a saturating intensity. Here, the IRs under different operating conditions are calculated from the forward and backward transmission coefficients (i.e., $T_{\text{F}}$ and $T_{\text{B}}$) with  
\begin{equation}
\text{IR}=10\log_{10}\left[\dfrac{\max\left\{T_\text{F},T_\text{B}\right\}}{\min\left\{T_\text{F},T_\text{B}\right\}}\right].
\label{IR}
\end{equation} 
The operations of two $\mathcal{T}$-symmetric waveguide variants in terms of nonlinearity-induced optical isolations are illustrated in Figs. \ref{fig6} and \ref{fig7}. Figure \ref{fig6} shows prototype isolation schemes for both the variants with the one-way transfer of selective modes at an optimized nonlinearity level ($N_l$), where Fig. \ref{fig7} illustrates how we optimize such a specific $N_l$.  

In Fig. \ref{fig6}(a), we show the one-way propagation of modes through WG$_\text{A}$ (which hosts a dynamically encircled EP) with Kerr-type nonlinearity in the spatial index distribution. We judiciously optimize the nonlinearity level at $N_l=6.75\%$. Here, we observe that the waveguide is active for the encirclement in the CW direction, where both quasiguided modes are fully transmitted from $z=0$ to $z=l$. Moreover, the adiabatic and nonadiabatic relations [from \eqref{ncf}] for this specific encirclement condition allow the asymmetric conversions $\left\{\Psi_{\text{F}},\Psi_{\text{H}}\right\}\rightarrow\Psi_{\text{H}}$, which results in delivery of the dominating $\Psi_{\text{H}}$ at $z=l$ of WG$_\text{A}$. Meanwhile, for the consideration of the dynamical EP encirclement in the CCW direction, we also observe that almost no light is transmitted from $z=l$ to $z=0$, which is shown in Fig. \ref{fig6}(b) via a relative intensity difference. Figure \ref{fig6}(b) schematically shows the prototype isolation scheme achieved using WG$_\text{A}$ along with one of the output (O/P) field intensities at both $z=l$ and $z=0$ (i.e., for the forward and backward transmissions, respectively, with the inputs as already shown in Fig. \ref{fig1}(b); as both modes are converted into a particular dominating mode for propagation in a specific direction, we obtain almost similar output intensities  at a particular output-end, and hence we show only one of two output field-intensities for each of the propagation directions). Here, the dotted blue curve represents the normalized output field intensity ($\Psi_{\text{H}}$) at $z=l$, while considering the forward propagation ($z:0\rightarrow l$). However, during the backward propagation ($z:l\rightarrow 0$), the dotted red curve shows the output field intensity ($\Psi_{\text{H}}$) at $z=0$, which is relative with respect to the output at $z=l$ obtained during the forward propagation (the relative output is considered to indicate the intensity difference while considering the propagation in two opposite directions). Here, output intensity at $z=0$ decreases almost 98.6\% (during the backward propagation) in comparison to the output at $z=l$ (during the forward propagation). Two outputs at $z=0$ and $z=l$ perfectly imply the prototype isolation scheme of WG$_\text{A}$, which passes $\Psi_{\text{H}}$ in the forward direction and blocks $\Psi_{\text{F}}$ in the backward direction. We calculate the IR using \eqref{IR}, where a maximum of the IR of 18.6 dB is achieved.

In Fig. \ref{fig6}(c), we investigate a prototype isolation scheme based on WG$_\text{T}$ in the presence of Kerr-type nonlinearity with $N_l=6.75\%$ (same as considered for WG$_\text{A}$). Here, we observe that the waveguide is surprisingly active in the backward direction ($z:l\rightarrow0$) which is associated with the CW dynamical encirclement scheme around the EP*. The waveguide passes the dominating $\Psi_{\text{F}}$ [based on the corresponding nonadiabatic correction factors from \eqref{ncf}] with the asymmetric conversions $\left\{\Psi_{\text{F}},\Psi_{\text{H}}\right\}\rightarrow\Psi_{\text{F}}$, as can be observed via the associated beam propagation results. The light becomes blocked in the forward direction, which is associated with the CCW dynamical encirclement process around the EP*. The prototype isolation scheme along with the output (O/P) field intensities for WG$_\text{T}$ are shown in Fig. \ref{fig6}(d). From the normalized output intensity at $z=0$ ($\Psi_{\text{F}}$; during the backward propagation) and relative output intensity at $z=l$ ($\Psi_{\text{H}}$; during the forward propagation; relative with respect to the output $\Psi_{\text{F}}$ at $z=0$), it is clearly evident that the intensity decreases $\approx93.3\%$ during the forward propagation through WG$_\text{T}$. Hence, WG$_\text{T}$ allows $\Psi_{\text{F}}$ to pass in the backward direction, however, blocks $\Psi_{\text{H}}$ in the forward direction, where we achieve a maximum of the IR of 11.75 dB.

Hence, at a particular nonlinearity level, both $\mathcal{T}$-symmetric waveguide variants behave as isolators, which allow the nonreciprocal transmission of two different modes in opposite directions. For a particular variant, a breakdown of the inversion symmetry in the length-depended gain-loss variation occurs in two opposite directions, where the tailored nonlinearity induces nonreciprocity. Hence, the intensity of the incoming waves becomes completely attenuated in a particular direction, despite being transmitted fully in the opposite direction. Here, a correlation between the nonreciprocal transmissions to two different allowed modes in two waveguide variants is dictated by the nonadiabatic corrections around EP and EP*.

Such an exclusive nonreciprocal transmission of selective modes mainly relies on the interplay between dynamical gain-loss variation (active components) and the tailored local nonlinearity in the spatial index distribution (passive components). During the propagation of light around an EP in the presence of nonlinearity, the complex $\beta$-values of the supported modes become affected significantly. The EP-induced interactions are led by the variations of both Re($\beta$) (modal confinement) and Im($\beta$) (decay rates), where the incorporation of nonlinearity directly influences Re($\beta$). Now, the mode confinement factors enhance with an increasing amount of nonlinearity, which results in the simultaneous reduction of the associated decay rates. Hence, the onset of nonlinearity modifies the gain-loss parameter space concerning the location of the EP (or EP*), and accordingly the relative-gain factor [$\Delta\gamma^{\text{ad}}_{\text{F,H}}$; associated with \eqref{ncf}] between the interacting modes is affected significantly during the evolution of modes following the dynamical EP-encirclement scheme. Based on such an interplay, the relative intensity difference at two opposite output-ends varies for different nonlinearity amounts, which can be understood from the variation of the IR concerning the nonlinearity level ($N_l$), as shown in Fig. \ref{fig7}(a).

The IR initially increases with an increasing Kerr-type nonlinearity level and takes a maximum value of 18.6 dB for WG$_\text{A}$ and 11.75 dB for WG$_\text{T}$ at a certain threshold nonlinearity-level of $N_l=6.75\%$ [as shown in Fig. \ref{fig7}(a)]. Here, the difference in the IR for two waveguide variants at a particular $N_l$ can be observed, which occurs due to a different gain-loss profile (exactly opposite; based on $\mathcal{T}$-symmetry) as can be seen in Fig. \ref{fig3}(b). The operation of WG$_\text{A}$ is mainly dominated by loss, whereas WG$_\text{T}$ operates with an overall higher amount of gain. Hence, WG$_\text{A}$ is able to induce a comparably higher output intensity difference for the light propagation in two opposite directions. An additional gain-amplification in WG$_\text{T}$ might reduce such intensity difference between two outputs, which results in achieving a lower IR for WG$_\text{T}$ in comparison to WG$_\text{A}$ at a particular $N_l$. However, we interestingly observe that both waveguide variants achieve their highest IR at a specific $N_l=6.75\%$, which affirms their operational correlation based on the chiral behavior of two conjugate EPs. It is further noticeable that while increasing $N_l$ more than 6.75\%, the IR decreases gradually [as shown in Fig. \ref{fig7}(a)] for both variants. Such a decrease of the IR after a certain threshold is mainly due to the abrupt effect of nonlinearity on the encirclement loop that affects the location of the EP significantly (i.e., the EP might come closer to the boundary of the modified loop in the parameter space due to a higher amount of nonlinearity). Here, judicious care should be taken to optimize the $N_l$, as a higher nonlinearity after a certain limit may exclude the EP from the parametric loop, for which the overall observation might be intangible. However, there is a sufficient scope of scalability to investigate the device operation for different amounts of nonlinearities within a broad range. The characteristic curve shown in Fig. \ref{fig7}(a) defines the process to choose an optimized nonlinearity amount, from where we set $N_l=6.75\%$ to obtain the beam propagation results in Fig. \ref{fig6}.

Then, instead of local Kerr-type nonlinearity, we introduce saturable nonlinearity in the spatial index distribution to investigate the nonreciprocal transmission through WG$_\text{A}$ and WG$_\text{T}$. The saturable nonlinearity is considered with a chosen saturating intensity [$I_s$; as per \eqref{saturable}] based on the materials of the background waveguide. For Kerr-type nonlinearity, a nonlinear interaction of light in the optical medium gradually increases with an increasing signal intensity, which might ensemble instability in the output signals after a certain limit. In this context, the consideration of the saturable intensity in the associated nonlinear interactions can potentially stabilize the output signals, where we can observe a higher intensity difference at two output ends for propagation in the opposite directions. Hence, we optimize the saturable nonlinearity level at 7.5\% from the characteristic dependence of the IR on $N_l$, as shown in Fig. \ref{fig7}(b). Here, we observe an exactly similar nonreciprocal response of both WG$_\text{A}$ and WG$_\text{T}$, as we have seen for the choice of Kerr-type nonlinearity in Fig. \ref{fig6}. WG$_\text{A}$ allows the nonreciprocal transmission of $\Psi_{\text{H}}$ in the forward direction, whereas isolates $\Psi_{\text{F}}$ in the backward direction. The field intensity decreases $\approx99.96\%$ during the backward propagation, where we achieve a maximum of the IR of 34.6 dB. On the other hand, we achieve a maximum of 18.6 dB IR for WG$_\text{T}$, which allows $\Psi_{\text{F}}$ to transmit along the backward direction and isolates $\Psi_{\text{H}}$ in the forward direction with almost 98.7\% reduction of the signal intensity.        

\section{Summary}

In summary, a significant stride in understanding and utilizing the concept of conjugate EPs has been made in the context of a correlative nonreciprocal light transmission process. Besides hosting the dynamical encirclements of two conjugate EPs in two $\mathcal{T}$-symmetric variants using the framework of a planar gain-loss assisted waveguide, a comprehensive all-optical platform has been developed based on the onset of nonlinearity along with the encirclement scheme to achieve nonreciprocal transmission of selective modes with a specific chiral correlation. We have established that two $\mathcal{T}$-symmetric waveguide variants, hosting two conjugate EPs, are characterized by their ability to behave as isolators enabling nonreciprocal transmission of selective modes in opposite directions. Here, they allow active transmission of two different dominant modes in opposite directions, whereas block light from passing in their respective reverse directions. We have investigated the effect of both Kerr-type and saturable nonlinearities on achieving nonreciprocity, where we have observed that the introduction of saturable nonlinearity can induce a comparably higher nonreciprocal effect. We have achieved a huge isolation ratio, even up to 34.6 dB under a specific operating condition. The intricate interplay of the dynamical gain-loss parameter space around the conjugate EPs in the presence of different types of nonlinearities has been discussed in detail to understand such unconventional chiral light dynamics. The insights and implementations of our approach harnessing the fascinating features of conjugate EPs in nonlinear optical systems would unlock a new avenue with exciting possibilities for boosting the development of various nonreciprocal components, such as optical isolators and circulators, for integrated (on-chip) photonic applications in next-generation communication networks and quantum information processing.              

\section*{acknowledgments}
A.L. and A.M. acknowledge the financial support from the Maestro Grant (No. DEC-2019/34/A/ST2/00081) of the Polish National Science Center (NCN). A.L. also acknowledges the support from the National Postdoctoral Fellowship Grant (No. PDF/2021/001322) of the Science and Engineering Research Board (SERB), India.

\section*{Appendix}\label{appendix}
The occurrence of conjugate EPs in any physical system can be understood as a mathematical problem by constructing an analogous $2\times2$ non-Hermitian Hamiltonian given by
\begin{equation}
\mathcal{H}(\lambda)=\mathcal{H}_0+\lambda \mathcal{H}_p=\left(\begin{array}{cc}\beta_1 & 0 \\0 & \beta_2 \end {array}\right)+\lambda\left(\begin{array}{cc}\kappa_1 & \gamma_1 \\\gamma_2 & \kappa_2\end {array}\right).\tag{A.1}
\label{ch2_H}  
\end{equation}
Here, a passive Hamiltonian $\mathcal{H}_0$, consisting of two passive eigenvalues $\beta_j\,(j=1,2)$, is subjected by a perturbation $\mathcal{H}_p$, which is dependent on some coupling parameters $\kappa_j$ and $\gamma_j\,(j=1,2)$ with a perturbation strength $\lambda$. 

A trivial case can be considered with real-valued $\beta_j$, $\kappa_j$, and $\lambda$ along with $\gamma_j=0$, for which the effective Hamiltonian $\mathcal{H}$ behaves as a Hermitian system and possesses two distinct eigenvalues: $\mathcal{E}_j(\lambda)=\beta_j+\lambda\,\kappa_j\,(j=1,2)$. Here, a conventional degeneracy occurs at $\lambda=-(\beta_1-\beta_2)/(\kappa_1-\kappa_2)$. Now, to ensure the system to be non-Hermitian, all the elements in $\mathcal{H}_p$ might be chosen as non-zero with a complex $\lambda$, where $[H_0,H_p]\ne0$. The operation of our designed dual mode waveguide based optical system can be understood based on such a non-Hermitian Hamiltonian. Here, $\beta_1$ and $\beta_2$ represent two real propagation constants. The complex $\lambda$ defines the overall non-Hermitian elements based on gain-loss parameters $\kappa_j$ and $\gamma_j\,(j=1,2)$, where $\kappa_j$ can be appeared as individual modal decay rates, whereas $\gamma_j$ can be considered as introduced gain-loss elements.    

The eigenvalues of $\mathcal{H}$ can generically be written as, 
\begin{equation}
\mathcal{E}_{1,2}(\lambda)=\frac{\beta_1+\beta_2+\lambda\left(\kappa_1+\kappa_2\right)}{2}\pm R;\tag{A.2}
\label{ch2_E} 
\end{equation} 
where,
\begin{align}
R=&\left[\left(\frac{\beta_1-\beta_2}{2}\right)^2+\lambda^2\left\{\left(\frac{\kappa_1-\kappa_2}{2}\right)^2+\gamma_1\gamma_2\right\}\right.\notag\\
&\left.+\frac{\lambda}{2}\left(\beta_1-\beta_2\right)\left(\kappa_1-\kappa_2\right)\right]^{1/2}.\tag{A.3}
\label{ch2_c} 
\end{align}
Owing to the coupling invoked by finite $\gamma_j\,(j=1,2)$, two levels $\mathcal{E}_1$ and $\mathcal{E}_2$ exhibit avoided resonance crossing (ARC; i.e., two levels do not cross but avoid each other) type interactions with a continuous variation of $\lambda$. While exhibiting ARCs, the two levels coalesce at two critical values of $\lambda$, which represent a complex conjugate pair of EPs in the complex $\lambda$-plane. These two singularities can be obtained in the complex $\lambda$-plane by setting $R=0$, which are given by 
\begin{equation}
\lambda_s^{\pm}=-\frac{(\beta_1-\beta_2)}{(\kappa_1-\kappa_2)\mp2i\sqrt{\gamma_1\gamma_2}} \tag{A.4}
\end{equation}
The connection between two conjugate EPs can be understood by rewriting $R$ in terms of $\lambda_s^+$ and $\lambda_s^-$ as
\begin{equation}
R=\left[\left(\dfrac{\lambda-\lambda_s^+}{2}\right)\left\{\left(\frac{\kappa_1-\kappa_2}{2}\right)^2+\gamma_1\gamma_2\right\}\left(\dfrac{\lambda-\lambda_s^-}{2}\right)\right]^{1/2}.\tag{A.5}
\label{ch2_E1} 
\end{equation}
Hence, the coupled levels are specified by the value of $\sqrt{\lambda-\lambda_s^+}$ and $\sqrt{\lambda-\lambda_s^-}$ on two different Riemann surfaces. The critical eigenvalues at two conjugate EPs (i.e., at $\lambda_s^+$ and $\lambda_s^-$) are given by
\begin{equation}
\mathcal{E}_s(\lambda_s^{\pm})=\dfrac{(\kappa_1\beta_2-\kappa_2\beta_1)\mp i\sqrt{\gamma_1\gamma_2}(\beta_1+\beta_2)}{(\kappa_1-\kappa_2)\mp 2i\sqrt{\gamma_1\gamma_2}}\tag{A.6}
\label{ch2_el}
\end{equation}
Now, an EP is associated with the occurrence of only one independent eigenvector, unlike two orthogonal eigenvectors at a trivial Hermitian degeneracy. Thus, using the bi-orthogonal norm for a non-Hermitian Hamiltonian, two right-hand eigenvectors at two conjugate EPs (one for each of the EPs) can be written as (approximated up to a factor)
\begin{subequations}
	\begin{align}
	|\Psi_s^+\rangle&=\left(\begin{array}{c}\dfrac{+i\gamma_1}{\sqrt{\gamma_1\gamma_2}}\\1\end{array}\right)\qquad\textnormal{for}\quad\lambda=\lambda_s^+,\tag{A.7a}\\
	|\Psi_s^-\rangle&=\left(\begin{array}{c}\dfrac{-i\gamma_1}{\sqrt{\gamma_1\gamma_2}}\\1\end{array}\right)\qquad\textnormal{for}\quad\lambda=\lambda_s^-;\tag{A.7b}
	\end{align}
	\label{ch2_rhe}
\end{subequations}%
with the associated left-hand eigenvectors
\begin{subequations}
	\begin{align}
	\langle\widetilde{\Psi}_s^+|&=\left(\begin{array}{cc}\dfrac{+i\gamma_2}{\sqrt{\gamma_1\gamma_2}}&1\end{array}\right)\qquad\textnormal{for}\quad\lambda=\lambda_s^+,\tag{A.8a}\\
	\langle\widetilde{\Psi}_s^-|&=\left(\begin{array}{cc}\dfrac{-i\gamma_2}{\sqrt{\gamma_1\gamma_2}}&1\end{array}\right)\qquad\textnormal{for}\quad\lambda=\lambda_s^-.\tag{A.8b}
	\end{align}
	\label{ch2_lhe}
\end{subequations}
From Eqs. (\ref{ch2_rhe}) and (\ref{ch2_lhe}), it is evident that 
\begin{equation}
\langle\widetilde{\Psi}_s^+|\Psi_s^+\rangle=0\qquad\textnormal{and}\qquad\langle\widetilde{\Psi}_s^-|\Psi_s^-\rangle=0,\tag{A.9}
\label{ch2_self}
\end{equation}
These conditions are referred to as the self-orthogonality that holds at both the conjugate EPs. The existence of only one self-orthogonal eigenvector reflects the fact that the Hamiltonian $\mathcal{H}(\lambda)$ becomes non-diagonalizable for both $\lambda=\lambda_s^+$ or $\lambda_s^-$, i.e., at the two conjugate EPs.

\bibliography{ref}

\end{document}